\documentclass[%
 reprint,
 amsmath,amssymb,
 aps,
]{revtex4-2}

\usepackage{graphicx}
\usepackage{dcolumn}
\usepackage{bm}
 \usepackage{gensymb}

\usepackage[usenames,dvipsnames]{xcolor}
\usepackage{ulem}

\begin{document}

\preprint{APS/123-QED}

\title{Multiscale geometrical and topological learning in the analysis of \\soft matter collective dynamics}

\author{Tetiana Orlova}
 \email{tetiana.orlova@ysu.am}
 \altaffiliation[Also at ]{Institute of Physics, Yerevan State University, 1 Alex Manoogian, Yerevan 0025, Armenia.}
\author{Malgosia Kaczmarek}%
 \email{mfk@soton.ac.uk}
\affiliation{%
 Physics and Astronomy, University of Southampton, Southampton SO17 1BJ, UK\\
}%

\author{Amaranta Membrillo Solis}
 \email{i.a.membrillosolis@qmul.ac.uk}
 \altaffiliation[Also at ]{School of Mathematical Sciences, Queen Mary University of London, London E1 4NS, UK.}
 \author{Tristan Madeleine}%
 \author{Giampaolo D’Alessandro}%
 \author{Jacek Brodzki}
\affiliation{
 Mathematical Sciences, University of Southampton, Southampton SO17 1BJ, UK\\
}%

\author{Hayley R. O. Sohn}
 \author{Ivan I. Smalyukh}
 \affiliation{%
 Department of Physics, University of Colorado, Boulder, CO, USA\\
 International Institute for Sustainability with Knotted Chiral Meta Matter (WPI-SKCM2), Hiroshima University, Higashi Hiroshima, Hiroshima, Japan\\
 Renewable and Sustainable Energy Institute, National Renewable Energy Laboratory and University of Colorado, Boulder, CO, USA\\
}%


\begin{abstract}
Understanding the behavior and evolution of a dynamical many-body system by analyzing patterns in their experimentally captured images is a promising method relevant for a variety of living and non-living self-assembled systems. The arrays of moving liquid crystal skyrmions studied here are a representative example of hierarchically organized materials that exhibit complex spatiotemporal dynamics driven by multiscale processes. 
Joint geometric and topological data analysis (TDA) offers a powerful framework for investigating such systems by capturing the underlying structure of the data at multiple scales. In the TDA approach, we introduce the $\Psi$-function, a robust numerical topological descriptor related to both the spatiotemporal changes in the size and shape of individual topological solitons and the emergence of regions with their different spatial organization. 
The geometric method based on the analysis of vector fields generated from images of skyrmion ensembles offers insights into the nonlinear physical mechanisms of the system's response to external stimuli and provides a basis for comparison with theoretical predictions. 
The methodology presented here is very general and can provide a characterization of system behavior both at the level of individual pattern-forming agents and as a whole, allowing one to relate the results of image data analysis to processes occurring in a physical, chemical, or biological system in the real world.
\end{abstract}

\maketitle


\section{\label{Intro}Introduction}

The formation of patterns is a well-known phenomenon observed in physical, chemical and biological systems as a consequence of non-linear dynamics leading to self-organisation of the system and its non-trivial spatio-temporal behavior \cite{Cross1993, Maini1997, Gollub1999, Knobloch2015, Li2020, DAlessandro1991a, Residori2005, Eber2005}. In fact, patterns can arise at different hierarchical levels as a result of the movement and interaction of a large number of multi-scale subsystems that constitute a complex system. Constituents range from molecular to macroscopic scale and include photons, atoms and molecules, organelles and living cells, particle clusters, bacteria, biological organisms flocking together, and stars forming a cosmic network. An analysis of the spatial structure and the temporal behavior of patterns can reveal the dynamics of processes that occur both at the macroscopic and microscopic levels of the underlying system.

Often complex system measurements are encoded in data point clouds, spanning from numerical data on structured grids, e.g. images, to networks and graphs, e.g. a gene-regulatory network. Topological data analysis (TDA) is a recent and rapidly growing field that provides new topological and geometric analytical tools to uncover the underlying features of complex systems from their measured data point clouds \cite{Chazal2021}. TDA is particularly powerful in extracting relevant topological and geometric features from complex data and provide valuable, multi-scale insight. A widely used TDA method to compute topological features is persistence homology which has been successfully applied in biology, medicine, chemistry, physics, and material science \cite{Meng2020, Saadatfar2017, Townsend2020, Suzuki2021, Masoomy2021, Leykam2021}. A new development in this research field involves introducing TDA into machine learning (ML) methods to exploit topological properties in ML pipelines or to use topological information to improve ML pipelines \cite{Hensel2021, Calcina2021, Leykam2023}.

Even though linking characteristics obtained from the persistent homology of complex systems to their chemical, physical and biological parameters is challenging, this has been attempted successfully using a wide range of experimental and simulated data.  Early studies have used persistent homology to analyse the equilibria and the periodic dynamics of the Rayleigh–Bénard convection and Kolmogorov flow, two model systems of spatio-temporal pattern formation away from equilibrium \cite{Cross1993, Bodenschatz2000, Kurtuldu2011, Kramar2016}. The global behavior of biological aggregations such as bird flocks, fish schools, and insect swarms have been quantified using Betti numbers, topological barcodes and different types of distance matrices \cite{Topaz2015}. The same tools have been used to characterise the dynamics of islands in a  confluent cellular monolayer spreading on an empty space surface, and to track and classify the evolving shapes of interfaces between two monolayers of different cells in an antagonistic migration assay \cite{Bonilla2020}. A similar persistence homology-based technique has been used to study the pattern-forming transition in cooling granular gases obtained by numerical simulations \cite{Yano2020} and in experimental studies of phase transition in nematic liquid crystals doped with plasmonic nanoparticles \cite{Solis2022}. Using a model example of various imperfect lattices, an attempt has been made to perform topological measurements for pattern-forming systems providing both roll and dot patterns, transitions between which can occur when a control parameter is changed \cite{Shipman2023}. Furthermore, TDA combined with ML techniques has been applied for automatic detection of critical transitions in microstructured materials during two distinct pattern-forming processes, such as the spinodal decomposition of a two-phase mixture and the formation of binary-alloy microstructures during physical vapor deposition of thin films \cite{Abram2022}. As another example, the unsupervised classification of persistence images has made it possible to automate the categorisation of multicellular spatial patterns, whose organisation is controlled by the efficiency of mutual cell adhesion \cite{Bhaskar2022}. 

Recently, several attempts have been made to apply TDA to patterning in real systems at several hierarchical levels. Here, perhaps the best example comes from the biological world, namely a study of simulated zebrafish skin patterns from an agent-based model that quantified pigment cell dynamics and global pattern attributes on a large scale using TDA, computational geometry and interpretable machine learning method \cite{McGuirl2020, Cleveland2022}. Another very interesting inverse analysis of topological data has also been shown, when a particular region of a persistent diagram was directly linked to an inhomogeneous area of simulated transmission electron microscopy images of amorphous and liquid states of matter \cite{Uesugi2022}.

In the present study, we demonstrate the ability of TDA to reveal the periodic behavior of complex, organised patterning agents and identify global attractor-like dynamics. This follows earlier results on persistent homology successfully revealing the cause of the magnetization reversal process on the original microscopic magnetic domain structure \cite{Kunii2022} and the mechanisms of formation dynamics of magnetic domain patterns. 
In particular, we apply TDA to experimental data from a soft matter system, consisting of electrically powered dynamic ensembles of three-dimensional twisted structures in chiral nematic liquid crystals.\cite{Sohn2019, Sohn2020} Such 3D structures are characterised by skyrmion-like configurations of the liquid crystal (LC) director field. Skyrmions were originally identified in the magnetization textures of chiral magnets \cite{Muhlbauer2009}, introduced in particle physics \cite{Skyrme1961} and then in chiral liquid crystals \cite{Ackerman2014, Duzgun2018, Foster2019}, magnetic colloids \cite{Zhang2015}, evanescent 
electromagnetic fields \cite{Tsesses2018} and light \cite{Sugic2021, Mototake2022}. The choice of liquid crystals as a test-bed for our method is dictated by the ease of their manipulation, their long-term stability of supporting topological structures at room temperature, the easy visualisation of skyrmions, and their high responsiveness to applied external fields, ensuring transitions between topologically protected metastable states.

We introduce a new topological characteristic, the $\Psi$-function, that allows us to reliably detect periodic changes in the size of pattern-forming agents, as opposed to the algebraic norm or structural heterogeneity that was previously proposed to track temporal evolution of soft matter systems \cite{Solis2022}. At the same time, we show that the evolution of a dynamic and complex hierarchical system can be examined in general terms by computing distance matrices between different ensemble states. The behaviour of the system as a whole can be analysed in more detail by modeling the time-dependent imaging data as discrete vector fields, and then applying multidimensional scaling to it, an unsupervised learning method similar to principal component analysis but suitable for the analysis of large images.

The paper is structured as follows. In the next section Section~\ref{Methods} we present our experimental and theoretical tools, including the $\Psi$ function. 
This section covers in some detail the core mathematical definitions and derivations. However, the following experimental sections are self-contained so can be followed, in the first instance, without the mathematical details.  

Section~\ref{R_and_D} is the core of the paper. Section~\ref{Time-evolving} illustrates the general principles of the two analysis approaches taken in this paper, evolutionary and topological. The first is the subject of Section~\ref{ST_evol}: here we use distance matrices to create low dimensional embedding spaces that allow the visualisation of the global skyrmion structure dynamics. The second is used in Section~\ref{PerStruct} to extract the fine scale harmonic and anharmonic behaviour wich is a signature of the nonlinearity of the skyrmion dynamics. The conlusions summarise this paper and discuss possible applications to other soft-matter fields. Additionally, the paper contains a Supplemental Material (SM). 

\section{\label{Methods}Methods}

\subsection{\label{Methods41}Liquid crystal samples for the formation of topological solitonic structures} 

Localized topological field configurations such as skyrmions, hopfions, torons, twistions and some others have recently been found in magnetic systems, liquid crystals and light beams \cite{Wu2022,Li2023,Yang2025}. They represent field configurations with a non-trivial global topological structure, i.e. they cannot be transformed into a homogeneous field by continuous changes. These structures also correspond to a local minimum of the field energy and are thus metastable structures, the transformation or erasure of which requires an external influence with energy input. In liquid crystals, the most well-known topological solitonic structure is the toron, also known as the cholesteric spherulite or cholesteric bubble \cite{Pieransky}. Its basic element is a double-twist cylinder looped on itself, accompanied by  hyperbolic point defects above and below the equatorial torus plane to match the locally twisted LC director field with the surrounding uniform unwound state \cite{Smalyukh2010}. In general, a wide variety of localized topological structures can be obtained in thin layers of chiral nematic liquid crystals (CLCs) under conditions where the geometry of the LC sample, combined with strong perpendicular anchoring conditions on confining substrates, suppresses the winding of the cholesteric helix \cite{Ackerman2017new2}.

When preparing samples for our experiments, a commercially available nematic mixture ZLI2806 (EM Chemicals) was doped with the right-handed chiral additive CB-15 (EM Chemicals). The weight fraction used for the chiral dopant was chosen as C$_{dopant}$= 1/($\xi$ ·p), as needed to define the helicoidal pitch p of the subsequent chiral LC mixture to be of desired value, where $\xi$~=~+5.9~$\mu$m$^{-1}$ is the helical twisting power of the chiral dopant in the particular nematic host we use. The CLC mixture was additionally mixed with 0.1~wt\% of cationic surfactant Hexadecyltrimethylammonium bromide (CTAB, purchased from Sigma-Aldrich) to allow spontaneous generation of torons by means of relaxation from electrohydrodynamic instability, as described below. The samples were prepared by sandwiching the mixtures between indium tin oxide (ITO)-coated glass substrates.
Strong perpendicular boundary conditions were set for the CLC director by treating the glass substrates with polyimide SE1211 coatings (Nissan Chemical). The treatment was implemented by spin coating the ITO sides of glass substrates at 2700 rpm for 30 s, followed by a 5 min pre-bake at 90{\degree}C and a 1 h bake at 180{\degree}C. 

\subsection{\label{Methods42}Manipulation of an ensemble of localised twisted structures} 

\begin{figure}[h]
\includegraphics[width=\columnwidth]{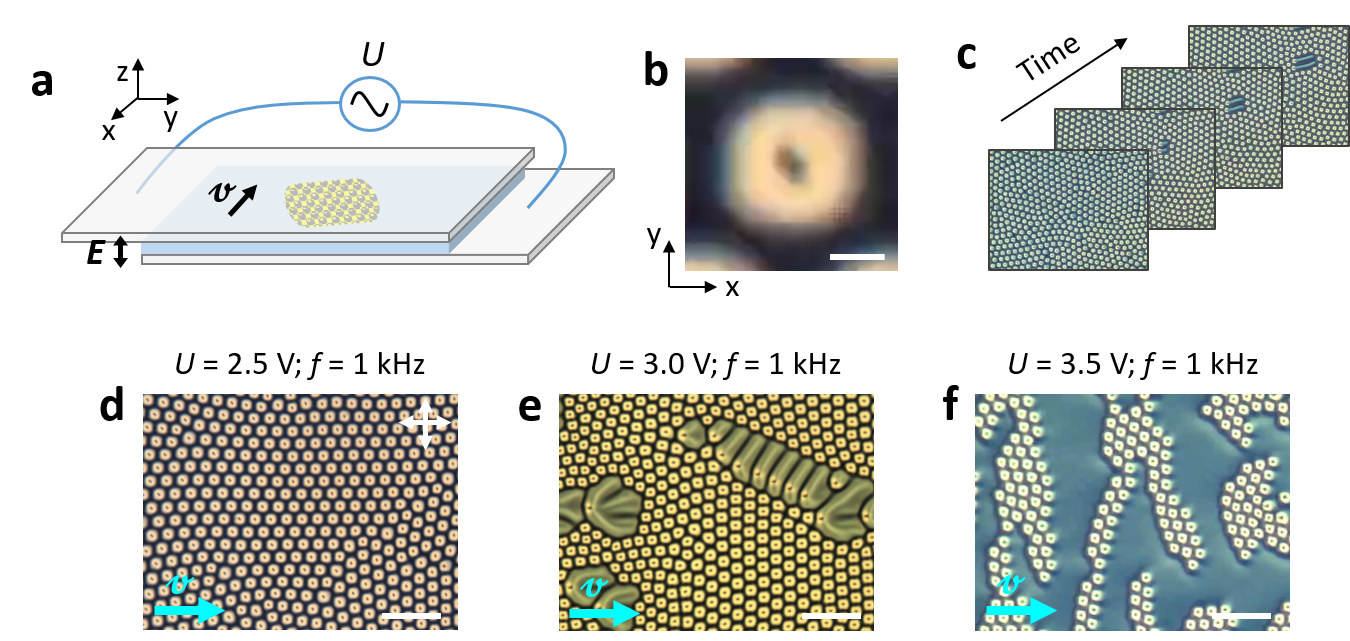}
\caption{Evolutionary and topological approaches to the analysis of complex soft dynamic systems. \textbf{a,}~Schematic of a chiral nematic LC sample under applied electrical voltage. \textbf{b,} Optical image of a single localized structure, representing its well-known topology with the spatial torus-like director field organization.\cite{Smalyukh2010} \textbf{c,} An example set of optical images demonstrating a time-evolving toron pseudo-crystallite with changes in the shape and position of localized structures. \textbf{d-f,} Examples of three dynamic systems, the ensembles of localized twisted structures, studied in this work. Experimental polarizing micrographs are taken after 1.5~min (\textbf{d}), 3~min (\textbf{e}) and 36.9~s (\textbf{f}) under an applied electric field.}\label{fig1}
\end{figure}

Although the topological solitons can appear spontaneously, in our experiments they were robustly generated by first inducing and then relaxing the electrohydrodynamic instability obtained at the applied AC voltage of $U$=20 V at the frequency $f$=2~Hz, forming spontaneously as energetically favorable structures after turning $U$ off (Fig.~\ref{fig1}a). This emergent robustness of the torons stems from the chiral CLC’s tendency to twist, which results in the formation of various twisted configurations. The particular twisted structures of torons that we study allow for relaxing the CLC's frustrated unwound state via formation of energetically favorable twist regions (Fig.~\ref{fig1}b). Furthermore, by manually switching on and off the electric voltage $U$ that induces the hydrodynamic instability 3-5 times in the course of a few seconds, one can control the number density as desired, up to tight packing of torons. The initial relative spatial positions of the torons are random, but crystallites slowly form due to the repulsive interactions at the high packing densities. The electric field needed to generate and control torons was applied across the samples using a custom made MATLAB-based voltage-driving program coupled with a data-acquisition board (NIDAQ-6363, National Instruments). Various electric driving schemes were used in order to morph the solitonic and power induced motions (Fig.~\ref{fig1}c). The macroscopically-supplied electrical energy was converted locally into solitonic motions that then exhibited various collective effects described in our study (Fig.~\ref{fig1} d-f). Optical videomicroscopy then allowed us to track the positions and collective organisations of the torons \cite{Senyuk2013,Ould-Moussa2013}.

\subsection{\label{Methods43}Distance matrices of image spaces} 

We use the spaces of vector fields over discrete measure spaces framework reported in \cite{membrillosolis2023} to analyse the video data. We provide here a brief description of the method to help with the discussion of the results.

We can think of an image as a vector field, $\mathcal X(\Lambda_{wh})$, defined over a grid of pixels $\Lambda_{wh}$ of width $w$ and height $h$. Let each pixel $s \in \Lambda_{wh}$ have equal weight or measure $\mu(s) = 1$. A grayscale image assigns a single real number (intensity) to each pixel, while an RGB image assigns a 3-dimensional vector (red, green, blue) to each pixel. Formally, this defines a function $\mathcal{X} : \Lambda_{wh} \to \mathbb{R}^d$, where $d = 1$ for grayscale and $d = 3$ for RGB.

To measure the difference between two images, we use what the $L^{p,q}$-norm of an image $\mathcal{X}$, given by:

\begin{equation}
\|\mathcal{X}\|_{L^{p,q}} := \left( \sum_{s \in S(\Lambda_{wh})} \|\mathcal{X}(s)\|_q^p \right)^{1/p}, \quad \text{if } p < \infty,
\end{equation}

and

\begin{equation}
\|\mathcal{X}\|_{L^{\infty,q}} := \max \{ \|\mathcal{X}(s)\|_q \},
\end{equation}

where $\|\cdot\|_q$ is the size of the intensity value of a pixel for a grayscale image, or the norm of the RGB vector, for an RGB image.

Given two images $\mathcal{X}, \mathcal{Y} : \Lambda_{wh} \to \mathbb{R}^d$, their $L^{p,q}$-distance is defined as:

\begin{equation}
\mathcal{L}^{p,q}(\mathcal{X}, \mathcal{Y}) = \left( \sum_{s \in S(\Lambda_{wh})} \|\mathcal{X}(s) - \mathcal{Y}(s)\|_q^p \right)^{1/p}.
\end{equation}

\begin{figure}[h]
\includegraphics[width=\columnwidth]{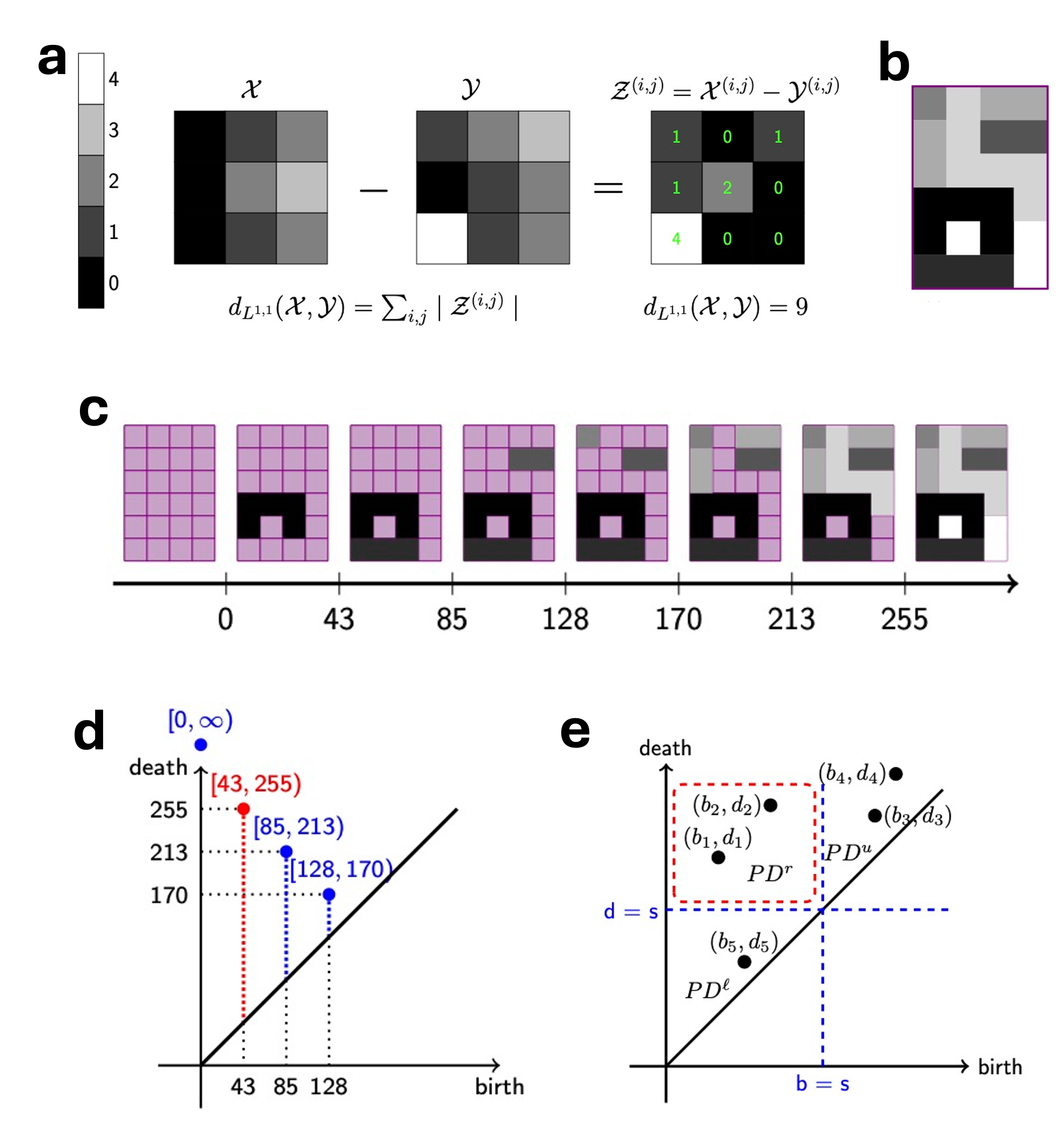}
  \caption{Geometric and topological methods for image analysis. \textbf{a,} Computation of the $L^{1,1}$ distance between two grayscale images. \textbf{b,} A grayscale image and \textbf{c,} the filtration by pixel intensity associated with it. As the pixel intensity increases, topological features such as clusters  of pixels and loops of pixels appear and disappear.  \textbf{d,} The Persistence diagram of the filtration is collection of points on the plane that records the information about the pixel intensity at which the 0-dimensional (blue) and one-dimensional  (red) topological features appear (birth), and the intensity value at which it disappears (death); \textbf{e,} Three separated regions $PD^r$, $PD^\ell$ and $PD^u$ at a fixed pixel intensity threshold $s$. The topological cycles $(b_1,d_1)$ and $(b_2,b_2)$ are the most stable at  pixel intensity $s$.}
    \label{fig:methods}
\end{figure}

This gives a flexible way to compare images, whether they are grayscale or RGB, by combining pixel differences across the entire image (Fig.~\ref{fig:methods}a). This method can also be used to measure the difference between the image gradients.

The RGB video frames obtained from Supplementary videos 1, 2 and 3 were transformed to grayscale. For each set of video frames, we computed the corresponding $L^{2,2}$-distance matrices. For these computations, we systematically selected every fifth video frame within the complete datasets for Supplementary videos 1 and 2, and every second video frame for Supplementary videos 3. The data processing and distance matrix computations were performed in Python.

\subsection{\label{Methods45}$L^{2,2}$-norms and $L^{2,2}$-distance matrices of video frames and gradients of video frames} 

The corresponding $L^{2,2}$-norms and distances matrices of the video frames were computed following the methodology described in the previous section.

The gradient of an image measures the rate and direction of change in pixel intensity, highlighting edges and regions of high contrast. In this work, gradients were computed using the \texttt{numpy.gradient} function in Python, which applies central differences for interior pixels and one-sided differences at the boundaries. For a 2D image $I$, the horizontal gradient $\frac{\partial\mathcal X}{\partial x}$ at pixel $(i, j)$ is approximated as:
\[
\frac{\partial \mathcal X}{\partial x} \approx \frac{\mathcal X[i, j+1] - \mathcal X[i, j-1]}{2}
\]
Similarly, the vertical gradient $\frac{\partial\mathcal X}{\partial y}$ is computed using values from adjacent rows. The $L^{2,2}$-gradient norm $|\nabla \mathcal X|$ is then calculated as:
\[
|\nabla \mathcal X| = \sqrt{\left(\frac{\partial \mathcal X}{\partial x}\right)^2 + \left(\frac{\partial \mathcal X}{\partial y}\right)^2}
\]
This norm emphasizes regions of significant local intensity change. The corresponding $L^{2,2}$-norms and distances matrices of gradients of video frames were also calculated in Python using the methodology described in the previous section. 

\subsection{\label{Methods44}Multidimensional scaling of distance matrices} 

We performed multidimensional scaling on the $L^{2,2}$-distance matrices to obtain Euclidean coordinates from the distance matrices associated with the videos J, N and C, which were obtained following the method described in section \ref{Methods43}. We performed multidimensional scaling using the MATLAB built-in function cmdscale.

\subsection{\label{Methods46}$k$-structural heterogeneity} 

In \cite{Solis2022}, structural heterogeneity was defined as a topological characteristic for soft matter systems, using imaging data. It is based on persistent homology, a data analytic tool used to get insight from the shape of the data. Here we generalise the notion of structural heterogeneity to consider two different types of topological features that might appear in imaging data. We recall some basic notion of persistent homology for grayscale images and the definition of structural heterogeneity and we introduce the generalised \textit{k-structural heterogeneity}. For a more detailed exposition of persistent homology, we refer the reader to \cite{otter2017,oudot2015}. 

A grayscale image can be analysed using persistent homology by examining how its structure changes across different light intensity levels (Fig. \ref{fig:methods}b-\ref{fig:methods}d). For each intensity threshold $i$ (ranging from 0 to 255), we create a filtration, i.e., a collection of simplified versions of the image that include only the pixels with intensity less than or equal to a fixed pixel intensity value $i$. As the threshold increases, new features appear and existing ones disappear. These features include isolated regions (connected components) and loops of pixels.

Each feature is recorded by the threshold value at which it appears (birth) and the value at which it disappears (death). These pairs of values are plotted to form a persistence diagram, which provides a summary of the image's topological structure across all intensity levels.

The difference between the death and birth values of a feature is called its persistence, and it reflects how long that feature remains present as the intensity threshold changes. By summing the persistence values of all loops (1-cycles) in the diagram, we obtain a measure of the image's structural complexity, referred to as its structural heterogeneity.

Given a grayscale image $\mathcal X$ with corresponding persistence diagram $PD(\mathcal  X)$, we define its $k$-structural heterogeneity, denoted $SH_k(\mathcal X)$, as the sum of the persistence values over all topological features or cycles $\alpha_k$ of dimension $k$ in  $PD(\mathcal X)$, with birth and death coordinates $(b_{\alpha_k},d_{\alpha_k})$: 

\begin{equation}\label{eq:kSH}
    SH_k(\mathcal X)=\underset{(b_{\alpha_k},d_{\alpha_k})\in PD(\mathcal X)}{\sum} d_{\alpha_k}-b_{\alpha_k}.
\end{equation}

Isolated regions define cycles of dimension zero or 0-cycles whereas loops of pixels define the cycles of dimension one or 1-cycles.
The open source TDA platform GUDHI  was used to compute the normalised persistence diagrams that include 0- and 1-cycles from grayscale video frames. The normalisation of persistence diagrams followed the procedure reported in \cite{Solis2022}.

\subsection{\label{Methods47}The $\Psi$-function} 

Extracting insightful topological information from noisy digital images can be challenging. Noise in images shifts the birth and death times of features in persistence diagrams, introducing many short-lived  features. While longer-lived features in a persistence diagram are more likely to represent real structures, short-lived ones are often artifacts of noise. To define a more robust numerical  topological descriptor than \eqref{eq:kSH}, we used the methodology  described in \cite[Section 3]{chung2018} to minimise the contribution of short-lived features, which are attributed to noise, while retaining the topological features that have a long lifespan. We define this optimised structural heterogeneity (OSH), denoted $\Psi$, next.

Let $PD_k(s)$ be the set of $k$-dimensional topological features that are ``alive'' at scale parameter $s$ in the persistence diagram $PD(\mathcal X)$, that is, features whose birth and death coordinates $(b, d)$ satisfy $b \leq s < d$. Then $|PD_k(s)|$ is the number of such features. Pictorially, $PD_k(s)$ corresponds to the set of points in the persistence diagram that lie to the left of the vertical line $b = s$ and above the horizontal line $d = s$. Let $PD(s) := \bigcup_k PD_k(s)$ be the union over all dimensions. Let $\Psi^N$ be defined as a normalised sum over all features alive at scale $s$, where each contribution is given by the product $(d - s)(s - b)$, with $b$ and $d$ denoting the birth and death of a feature, respectively:
\begin{equation}\label{eq:psifunction}
\Psi^N(s) = \frac{1}{|PD(s)| + 1} \sum_{(b,d) \in PD(s)} (d - s)(s - b).
\end{equation}

\noindent The product $(d-s)(s-b)$ in \eqref{eq:psifunction} becomes large only when $s$ lies near the middle of a feature’s lifespan that is, between its birth time $b$ and death time $d$. If $s$ is close to $b$ or $d$, one of the terms $(s-b)$ or $(d-s)$ becomes small, causing the product to shrink. 

Two correction factors need to be introduced to account for those topological features that were discarded from $PD(\mathcal X)$ in $\Psi^N(s)$. At threshold $s$, the persistence diagram splits into three regions (Fig. \ref{fig:methods}e): the rectangular region $PD^r$; the lower triangular region $PD^\ell$ consisting in all cycles whose death coordinates satisfy $d\leq s$; and the upper triangular region $PD^u$ consisting in all cycles whose birth coordinates satisfy $b > s$. Consider the  functions:

\begin{equation}\label{eq:psifunction}
\Psi^\ell(s) = \sum_{(b,d) \in PD^\ell(s)}\frac{s - d}{d-b}
\end{equation}
and
\begin{equation}\label{eq:psifunction}
\Psi^u(s) =  \sum_{(b,d) \in PD^u(s)}\frac{b-s}{d-b}.
\end{equation}

Give a grayscale image $\mathcal X$, the OSH of $\mathcal X$, denoted $\Psi$ is defined by 

\begin{equation}
\Psi(\mathcal X)=\underset{s\in[0,I]}{\text{argmax }}\Psi^\ell(s)\cdot\Psi^N(s)\cdot \Psi^u(s)
\end{equation}
where $[0,I]$ is the range of pixel intensity values. 
For each set of time-dependent images $\{\mathcal X(t)\}_{t\geq0}$, we can compute the corresponding time-dependent value $\Psi(\mathcal X(t))$ to keep track of the time-evolution of the associated dynamical system.

\section{\label{R_and_D}Results and Discussion}

\subsection{\label{Time-evolving}Time-evolving complex soft matter system}

In this section, we present three experimental systems used as model examples to demonstrate transformational topological approaches to the analysis of dynamic multi-scale soft-matter-based systems. As already mentioned, the pattern-forming agents are 3D localised twisted structures in liquid crystals, characterised by a complex spatial distribution of the LC director field. A variety of such particle-like structures spontaneously appear as elastic excitation in frustrated films of chiral nematics after the relaxation of electrohydrodynamic instabilities, provided that the helical pitch \textit{p} only slightly exceeds the thickness of the homeotropically oriented LC layer \textit{d} ($p \gtrsim d$) (see Methods~\ref{Methods41}) \cite{Kawachi1974, Ackerman2015}. These structures can also be created by optically induced reorientation of liquid crystal molecules when illuminating a frustrated chiral nematic film with a tightly focused laser beam of sufficient power \cite{Smalyukh2010, Chen2013, Loussert2014}. In this case, the spot size and the power of the light beam determine the type of elastic excitation formed. 

When an alternating electric field is applied to a dense-packed ensemble of particle-like soft structures (Fig.~\ref{fig1}a) (see Methods~\ref{Methods42}), shearing-like deformations of quasi-hexagonal lattices occur in a pseudo-crystallite, accompanied by the evolution of crystallite grain boundaries (Fig.~\ref{fig1}d) (Supplementary video 1). In addition, under certain parameters of the applied field, the transformation of pseudo-crystallites can be accompanied by the individual transformations of localised structures into cholesteric fingers (Fig.~\ref{fig1}e) (Supplementary video 2). If, however, an electric field is applied to a less dense ensemble of soft quasi-particles, each of the structures transforms the macroscopically supplied electrical energy in such a way that the entire ensemble exhibits collective motion in a direction that does not correlate with the direction of the electric field applied orthogonally to confining substrates (Fig.~\ref{fig1}f) (Supplementary video 3). This motion leads to the formation of dynamic chains and clusters of localized structures, so that in general the process resembles movement in complex self-assembling living systems, such as flocking of birds or fish \cite{Sohn2019}.

The analysis of dynamic ensembles of localised elastic excitations in a viscoelastic liquid-crystal medium is a complex problem. Even for a single skyrmion structure, studying the non-equilibrium behaviour is a challenge \cite{Ackerman2017, Duzgun2021}, while for dynamic toron pseudo-crystallites, so far only Voronoi reconstruction has been applied to reveal the complex movements of particles and defects within their lattices \cite{Sohn2020}. On the other hand, light passing through a three-dimensional LC structure placed between polarizers forms a distinct optical microscopic image, which is determined by the spatially inhomogeneous orientational distribution of molecules within the structure. Although it should be remembered that the optical image is a two-dimensional projection of the transmitted light and the reconstruction of the 3D director distribution is impossible without additional experimental data or numerical simulations, nevertheless, different localised excitations form optical images that are distinguishable from each other~\cite{Ackerman2017rew}, and an ensemble of localized excitations creates a patterned optical picture. Based on this, we introduce two methods for studying dynamic ensembles of localised structures: the first is an evolutionary approach that can be used to explore the general behavior of a nonequilibrium many-body system, even if individual structures undergo shape transformations. The second is a topological approach that can be applied to detect changes in the localised structures themselves that constitute a dynamic ensemble.

The first method implies that the time sequence of experimentally taken video frames visualizes changes in the state of the three-dimensional system (Fig.~\ref{fig1}c), and the numerically computed distance between frames (in other words, the degree of discrepancy between the intensity distributions of optical images) reflects the rate of these changes (Fig.~\ref{fig:methods}a). Then the reconstructed distance matrix between all video frames reveals the evolutionary path of the system. 

The second method is based on the fact that the intensity distribution is related to the topology of a localized structure (Fig.~\ref{fig1}b), and TDA reveals the structural features of the two-dimensional light pattern (Fig.~\ref{fig:methods}b-e). In each case of the dynamic ensembles presented in Fig.~\ref{fig1}d-f, consideration of the time dependencies of topological quantities allows us to detect changes in the 3D spatial organisation of the localized structures themselves, similarly to how we tracked structural changes in liquid crystal nanocomposites \cite{Solis2022}.

\subsection{\label{ST_evol}Spatio-temporal evolution of soft quasi-particle ensembles}

\begin{figure}[h]
\includegraphics[width=\columnwidth]{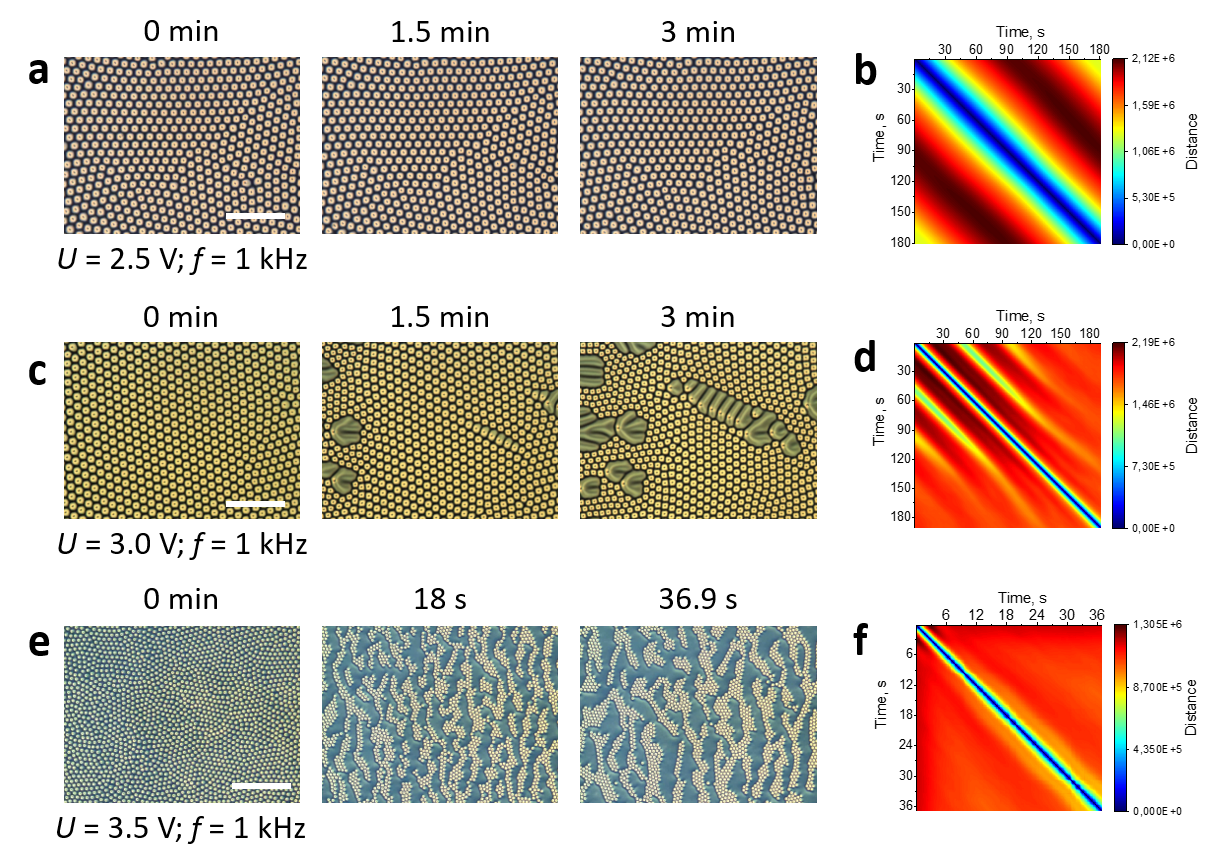}
\caption{Spatio-temporal evolution of different types of soft quasi-particle ensembles. \textbf{a,c,e,} Time evolution of three ensembles of localized structures under an applied electric field. \textbf{b,d,f,} The computed  image distance matrices for corresponding videos.}\label{fig2}
\end{figure}

As discussed earlier, for the evolutionary analysis, we consider the difference between frames of the same video to understand the difference between the physical states of a complex dynamical system. To quantify this difference, we computed the $L^{2,2}$-distance between all pairs of frames (see Methods~\ref{Methods43}). Inspection of the computed distance matrix (DM) allows us to reveal systems that preserve their overall spatial organisation over time evolution. In the case of shear-like deformation of the pseudo-crystallite (Fig.~\ref{fig2}a), the distance between any video frame and sequentially all the others first smoothly increases and then decreases again, demonstrating that the system tends to return to its initial configuration (Fig.~\ref{fig2}b). Simple visual observation of video frames shows that the grain boundary also drifts with time, but the DM turns out to be insensitive to this change. If the movement of a pseudo-crystallite is accompanied by a shape transformation of individual pattern-forming agents (Fig.~\ref{fig2}c), the system is only able to approach the initial state after moving away from it, and ultimately moves far from the initial configuration (Fig.~\ref{fig2}d). In addition, the much narrower blue diagonal in Fig.~\ref{fig2}d compared to Fig.~\ref{fig2}b indicates that in the second case the rate of change of the system is much higher, which is consistent with faster movement of the pseudo-crystallite (Supplementary video 1 and 2). In the case of self-assembly of moving clusters (Fig.~\ref{fig2}e), the distance between the states of the system  first increases remarkably quickly, which corresponds to the transition from pseudo-crystalline to cluster spatial organization, and then remains large and almost unchanged (Supplementary video 3), which is explained by displacement of clusters of different sizes and shapes over time (Fig.~\ref{fig2}f). The widening of the DM diagonal indicates that the cluster states of the system are closer to each other than the pseudo-crystalline and cluster states.

\begin{figure*}
\includegraphics[width=0.95\textwidth]{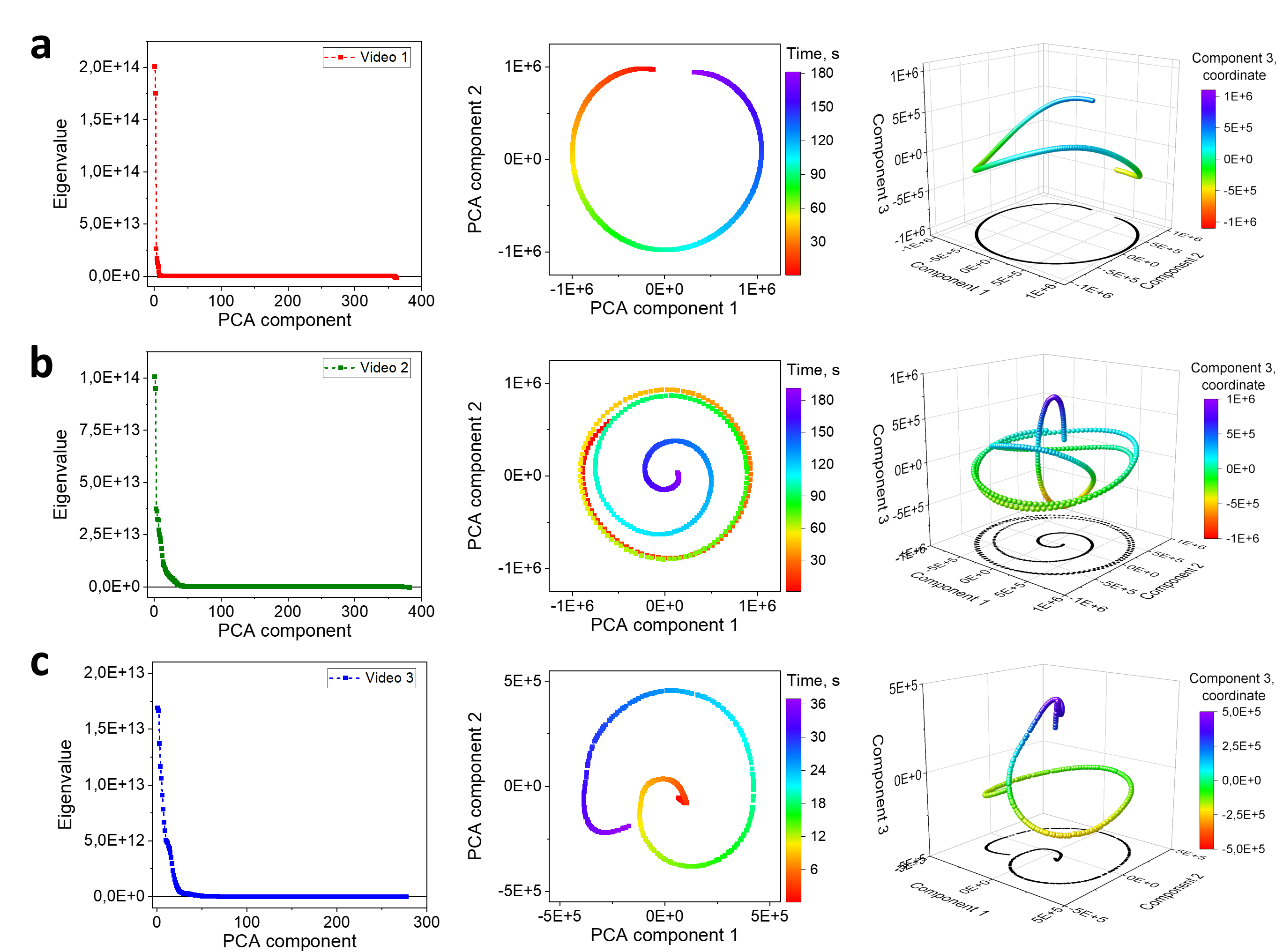}
\caption{Principal coordinate analysis applied to three different dynamic ensembles of localized twisted structures. \textbf{a,b,c,} Eigenvalues of the principal components and trajectories in the spaces of the first two and first three principal components for the moving ensembles of \textbf{a,} shape-persistent soft quasi-particles, \textbf{b,} quasi-particles experiencing shape transformation, \textbf{c,} clustering quasi-particles.}\label{fig3}
\end{figure*}

In general, DMs provide a qualitative understanding of the evolution of soft reconfigurable quasiparticle systems, but a more detailed analysis of the of the direction and end point of evolution is desirable. Since dynamic ensembles of skyrmions are multidimensional complex systems, it is reasonable to reduce the dimension of the video frame intensity data for further consideration, for example, using the principal component analysis method (PCA). Obviously, liquid-crystalline systems are essentially nonlinear, so one would expect that dimensionality reduction will also be nonlinear, but here we aim to consider how linear dimensionality reduction can be applied to such systems and what results can be obtained. Therefore, we applied multidimensional scaling for three videos, selected frames of which are presented in Fig.~\ref{fig2}a,c,e. Multidimensional scaling is an unsupervised method that allows one to obtain Euclidean coordinates of the $L^{2,2}$-distance matrices (see Methods~\ref{Methods44}). In the simplest case of densely packed and moving skyrmions (Fig.~\ref{fig3}a), the eigenvalue intensities show that the first two components are an order of magnitude stronger than the third one. In their two-dimensional space, an almost closed loop is formed. This corresponds to the situation when a densely packed hexagonal lattice of localized structures is preserved during its movement, and at a certain time, when the positions of the localized structures coincide with the initial ones, the system returns to its original state. When taking the third principal component into account, the 3D curve lies almost in a plane parallel to the plane of the first two components, which reflects the minor contribution of the third component to the system dynamics.

In the case of the second video (Fig.~\ref{fig3}b), a spiral-like converging trajectory is obtained in the space of the first and second components, leading to a seemingly attractive fixed point at the end of evolution when all localized structures are transformed into cholesteric fingers during their movement. However, from the eigenvalues it is clear that several more principal components are also significant. In 3D space of the first three components, a rather flat curve at the beginning of evolution indicates a minor contribution of the third component, but closer to the end of the video its contribution becomes remarkable. Thus, the vertical part of the 3D curve is formed representing the growth of cholesteric fingers, although in the projection onto the plane of the first two components, a converging curve resembling a fixed point attractor indeed appears.

It is difficult to isolate significant components in the case of the third video with skyrmion clustering (Fig.~\ref{fig3}c), but if we limit the analysis to the first three, then the contribution of the third component is strong at the beginning of the video, when clusters are formed and the translationally invariant configuration (TIC) between the skyrmions appears \cite{Ribiere1991,Smalyukh2005}. Then the contribution of the third component weakens, and the 3D curve becomes flatter, which corresponds to the movement of the clusters in the real-time video. The 2D projection shows a diverging spiral which we expect not to diverge much further since the clusters have already formed. Unfortunately, we are unable to verify this because the recorded video is too short. 
The temporal evolution of individual principal components (Fig.~S1 of the SM) also confirms the fact that the first and second components are of key importance when considering the movement of shape-persistent pattern-forming agents, while the third and even fourth components dominate in the case of their simultaneous shape transformation.  

Note that clustering can also be considered as a process of shape transformation of the pattern-forming agents. In the case of system shown in Fig.~\ref{fig2}e, this occurs by “gluing” skyrmions into their long chains surrounded by the TIC phase and forming a dynamic stripe-like pattern (Supplementary video 3), unlike the example shown in Fig.~\ref{fig2}c, where individual skyrmions themselves transform into stripe-like cholesteric fingers (Supplementary video 2).

To sum up, DMs are capable of capturing the similarity of states of time-evolving systems on a large size scale, without taking into account finer details and structural defects of lower dimensions (e.g., linear disclinations such as grain boundaries). Its coupling with PCA makes it possible to detect whether the movement of ensemble-forming structures is accompanied by their shape transformation or other, more complex processes, and when this occurs in time.

\subsection{\label{PerStruct}Periodic structural changes of soft quasi-particles}

\begin{figure}[h]
\includegraphics[width=\columnwidth]{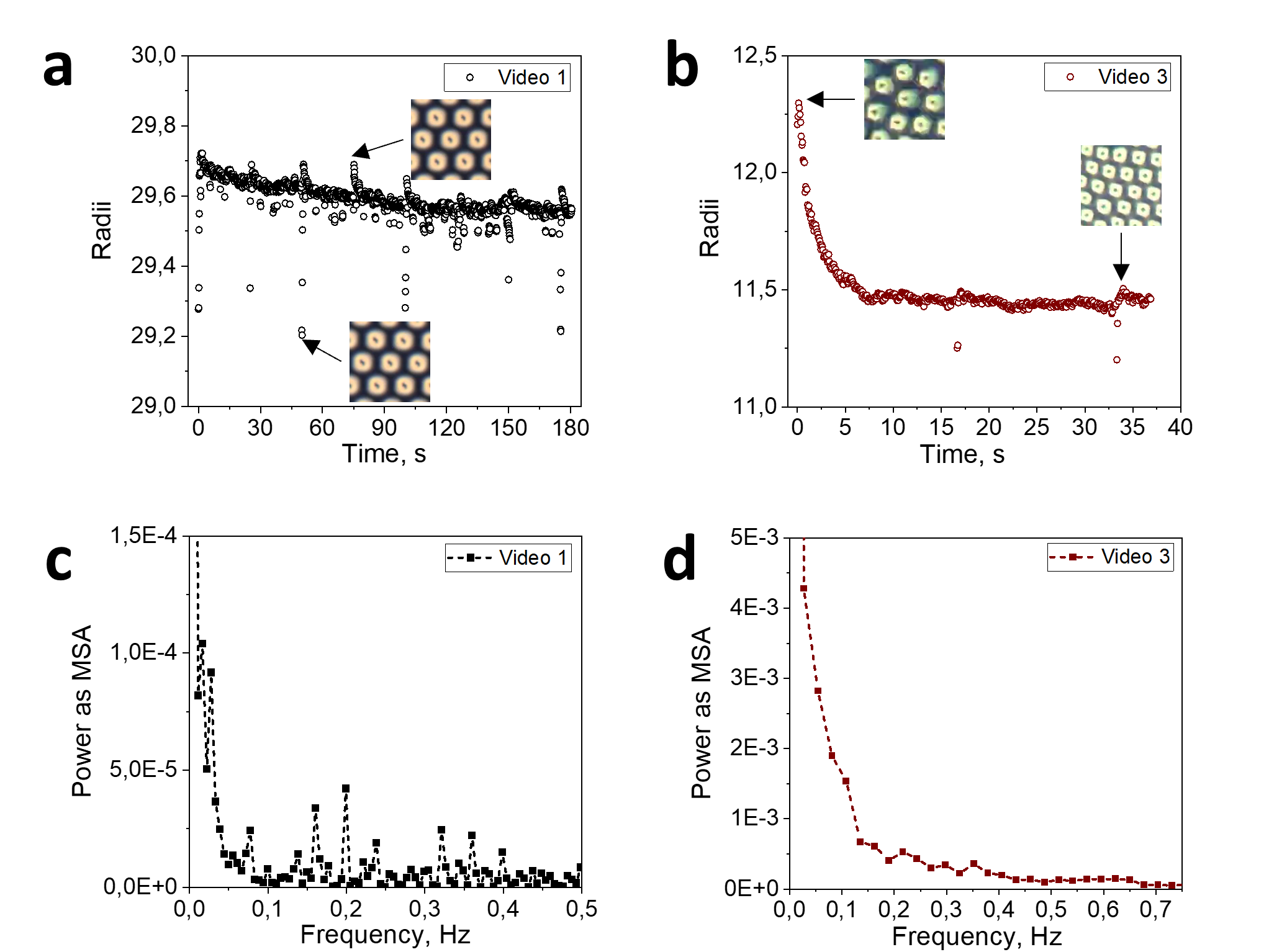}
\caption{Temporal changes in the average size of localized quasi-particles. \textbf{a,b,} Time evolution of averaged quasi-particle diameter, the pixel intensity threshold is set to 125 for \textbf{a} and 186 for \textbf{b}. \textbf{c,d,} Corresponding power spectra derived from FFT analysis. Only the first 10\% of the frequency domain is shown. The zero frequency is not shown due to its high power.}\label{fig4}
\end{figure}

Careful visual inspection of the video data prompted us to analyze in detail the size variation of localized structures, as we noticed their regular and consistent pulsation. In the case of a dynamic pseudocrystallite, the average size of localized structures changes in a small range of values, but clearly periodically (Fig.~\ref{fig4}a). In the case of moving self-organized clusters, the average size of quasiparticles decreases noticeably and then remains almost constant, however, from the two observed dips one can assume that in this system the variation period can be much longer (Fig.~\ref{fig4}b). Fast Fourier Transform (FFT) analysis applied to the first data set reveals a number of distinct frequencies (Fig.~\ref{fig4}c), while in the second case no periodicity in time can be clearly identified (Fig.~\ref{fig4}d). 

The observed variations in the size of localized structures are ensured by a change in the LC region size deviating from the orientation direction perpendicular to the sample substrates. The larger region and more molecules deviate from this direction, the higher the intensity of transmitted light through 90{\degree}-crossed polarizers, and the larger the observable structure size. The absence of noticeable distortions in the shape of the structures during their variations in size indicates that the director field orientational configuration inside changes symmetrically relative to the centers of the structures, while the topology most likely remains unchanged. Therefore, the algebraic $L^{2,2}$-norm of the intensity of video frames or even the gradients of video frames may be more suitable for analyzing the periodic behavior of this system (see Methods~\ref{Methods45}). However, only in the case of a dynamic pseudo-crystallite, FFT analysis of the norm of the video frame gradient gives a pronounced set of frequencies (Fig.~S2 a,b, of the SM), while in the case of shape-changing localized structures, the distinctness of frequencies is noticeably reduced (Fig.~S3 a,b, of the SM). In the case of clustering, as before, it is very difficult to distinguish any frequencies from the FFT data (Fig.~S4 a,b, of the SM). In an effort to reveal information about the periodic behavior of localized elastic structures and following the methodology for analyzing the time evolution of soft matter systems presented earlier in ~\cite{Solis2022}, we computed the time dynamics of structural heterogeneity of dimensions 0 and 1 (see Methods~\ref{Methods46}). Although for a dynamic pseudo-crystallite and a densely packed ensemble of shape-transforming structures, FFT analysis of the computed dependencies does not provide any additional information (Fig.~S2 c,d, and Fig.~S3 c,d, of the SM), in the case of clustering, some distinct frequencies can be detected (Fig.~S4 c,d, of the SM). Therefore, using TDA, it is possible to generate data that, at the next step of FFT analysis, reveals periodic changes of soft pattern-forming quasi-particles. 

\begin{figure*}
\includegraphics[width=0.95\textwidth]{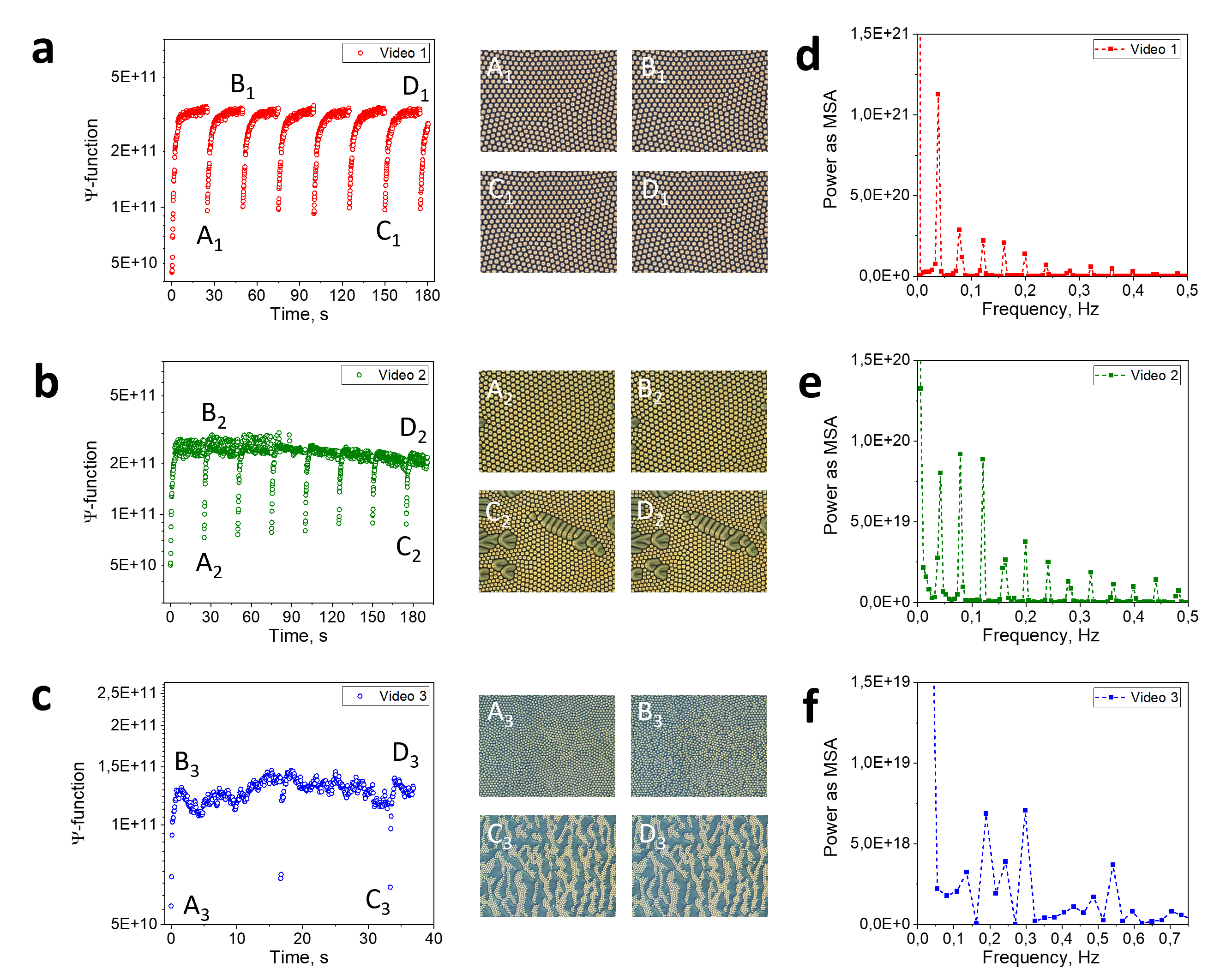}
\caption{Periodic behavior of a new topological descriptor, $\Psi$-function. \textbf{a,b,c,} Time evolution of topological $\Psi$-function computed for the case of \textbf{a,} moving pseudo-crystallite with shape-persistent localized structures, \textbf{b,} close-packed dynamic localized structures with shape transformation, \textbf{c,} dynamic clusters of localized structures. \textbf{d,e,f,} Corresponding power spectra derived from FFT analysis of the $\Psi$-functions. Only the first 10\% of the frequency domain is show in each case. The zero frequency is not shown due to its high power.}\label{fig5}
\end{figure*}

Aiming to obtain a distinct frequency spectrum revealing periodic changes in localized structures for different dynamic ensembles, we have constructed a new topological descriptor, the optimised structural heterogeneity (OSH) or $\Psi$-function. This numerical descriptor takes into account an optimised version of the 0 and 1-structural heterogeneity of the system (see Methods~\ref{Methods46}). In other words, the $\Psi$-function  quantifies different levels of organisation of molecules in a liquid crystal system, which give rise to the formation or disappearance of topological features, such as connected components or loops (see Methods~\ref{Methods47}). For the moving pseudo-crystallite, the $\Psi$-function changes periodically in time, but it can be noted that its initial dip in the first period differs from all others (Fig.~\ref{fig5}a). This may be due to the effect of electric field switching on the sizes of localized structures. A similar, but less pronounced effect can be observed in the case of a dynamic ensemble of shape-transforming structures (Fig.~\ref{fig5}b). In addition, one can see that in this case the amplitude of all other dips gradually decreases. We associate this with a decrease in the number of axisymmetric localized structures and the growth of cholesteric fingers, the behavior of which gives a different contribution to the variation of the transmitted light intensity during video recording and, ultimately, to the change in the $\Psi$-function. 

In the third case of dynamic self-assembled clusters, the effect of switching on the electric field is insignificant. The limited number of dips makes it difficult to explain the behavior of localized structures from visual inspection of the $\Psi$-function dependence (Fig.~\ref{fig5}c). Therefore, to achieve a more precise analysis, we again applied FFT and revealed that for the case of a dynamic pseudo-crystallite consisting of shape-persistent torons, the Fourier spectrum shows a set of frequencies that look like a fundamental frequency accompanied with harmonics (Fig.~\ref{fig5}d). However, in fact, the spectrum consists of the fundamental frequency and its second harmonic, while all other frequencies are slightly larger comparing to the exact harmonic values. These spectral components are interharmonics, i.e. at frequencies that are not integer multiple of the fundamental frequency. Note that the power of spectral components decreases sharply with increasing frequency, but not in agreement with an exponential law. In the case of dynamic and dense packed ensemble of shape-transforming localized structures, the first harmonic is observed at almost the same frequency as in Fig.~\ref{fig5}d, its additional harmonics are absent, but a spectrum of inter-harmonics is observed instead (Fig.~\ref{fig5}e). The contribution of the first several interharmonics to the spectrum is quite remarkable, since they have almost the same power as the first harmonic or even exceed it. However, we consider the first frequency to be fundamental, since it coincides well with the fundamental harmonic in the case of Fig.~\ref{fig5}d, and the physics of the processes occurring in these systems should be similar. With this definition, the spectrum shown in (Fig.~\ref{fig5}e) consists of a set of inter- and subharmonics.
Finally, in the case of toron forming dynamic clusters, a complex spectrum is obtained with two spectral components of similar power (Fig.~\ref{fig5}f). It is problematic to determine the fundamental frequency, but, in any case, the spectrum confirms that the processes occurring in this system are quite aperiodic.

The formation of sets of different harmonics may be related to the nature of the liquid crystal systems under study. As is known from signal analysis, the power values are proportional to the energy of each corresponding frequency. Harmonics arise from a pure sine wave processed by a nonlinear device, which in this case is the liquid crystal medium. Subharmonics show how much a signal deviates from a periodic shape, since only the harmonic family has the property that each member is also periodic with the fundamental period. Various subharmonics could appear when the driving frequency is over the fundamental frequency of the system due to the system nonlinear stiffness. Interharmonics indicate an increasing number of loads and nonlinearities in the system. Therefore, we come to the conclusion that the degree of nonlinear response increases from a dynamic quasi-crystallite consisting of torons to a moving ensemble of shape-transforming localized structures and especially to clustering torons. 
In the first case, the nonlinear response of a complex LC system leads to the emergence of a set of interharmonics that are close in frequency to the harmonics and provide a much smaller contribution compared to the fundamental frequency. In the second case, the nonlinear response becomes more complex, which could be associated with the formation of cholesteric fingers, and apparently leads to the fact that the contribution of interharmonics is even higher than the fundamental frequency. In the third case, the power spectrum could be explained by the dual response of the dynamic toron clusters and the observed large regions of the liquid crystalline TIC phase, the behavior of which, obviously, should be very different from the behavior of localized twisted structures.

\section{\label{Concl}Conclusions}

Dynamic complex patterns are observed in a wide variety of living and nonliving systems, typically driven the underlying physical, chemical or biological processes taking place. The evolution of such systems is determined by the behavior of both individual pattern-forming agents and the entire system as a whole. Thus, the in-depth investigation of their behavior requires the analysis at different hierarchical levels. In our study, we applied a number of geometric and topological methods to study the dynamics of patterns and their individual elements. 

As a particular case of a pattern-forming and multi-level system, we have analyzed the behavior of dynamic ensembles of localized twisted structures in chiral LCs, where the structures themselves could change their individual shape or vary the size while maintaining it. External electric field imparts a dynamic behavior, inducing a translational motion of localized structures, and changing their spatial organization, topology and size.

We found that geometric and topological data analysis are well suited to characterise the dynamics of such systems. The computed distance matrices between frames of recorded videos allow us to understand the evolution of the many-body system at a qualitative level, while consideration of principal component analysis data reveals not only the trajectories of the system in different phase spaces reflecting real changes, but also allows us to separate the evolution associated with the translational motion of pattern-forming structures from the movement accompanied by a change in their shape or their clustering. Furthermore, we introduced a new topological characteristic, the $\Psi$-function, which detects periodic processes in ensembles of pattern-forming agents with a constant topology and separates this scenario from other cases when, simultaneously with movement, the shape of individual structures changes or regions with significantly different spatial organization of the material appear. 

The presented approach could be useful in solving the problem of identifying and mapping domains in various living and abiotic systems. A prominent example of such a challenge is linking high-dimensional gene expression with three-dimensional cell morphology \cite{Way2022}, including the response of cell state to genetic and chemical perturbations \cite{Haghighi2022}, or to the form and function of living organisms \cite{Palande2023}. Understanding this relashionship remains a major problem in biomedicine \cite{Tang2025}, bioinformatics \cite{Ramos2025}, and general biology \cite{Ghavi2019,Skinner2023}. Other outstanding problems include the exploration of hierarchical bioinspired nanocomposites with embedded functionalities for dynamic and synergetic responses \cite{Nepal2023}, nanoparticle superclusters for light-harvesting nanomaterials in solar energy utilization \cite{Li2021}, and thin-film soft materials for green energy systems \cite{Shi2023}. Overall, our findings highlight that Topological Data Analysis is universal and should prove valuable for the study of a wide variety of dynamic self-assembled multi-level systems: from microtubules inside a cell, bacterial colonies and schools of fish to floating gel particles, reaction-diffusion waves, moving charged metal beads and convection cells, among many others.

\begin{acknowledgements}
This work was supported in part by the Leverhulme Trust (grant RPG-2019-055) and the EPSRC (grant EP/Y007484/1). TO acknowledges the support from the Higher Education and Science Committee of MESCS RA (Research Project no. 24IRF-1C003). 
\end{acknowledgements}

\bibliography{apssamp}

\end{document}